\documentclass[%
 reprint,
 amsmath,amssymb,
 aps,
]{revtex4-2}

\usepackage{graphicx}
\usepackage{bm}
\usepackage{hyperref}
\usepackage[capitalize]{cleveref}
\crefname{section}{Sec.}{Secs.}
\usepackage{physics}
\usepackage{xcolor}

\begin{document}

\title{Minimal quantum dot based Kitaev chain with only local superconducting proximity effect}

\author{William Samuelson}
\affiliation{%
Division of Solid State
Physics and NanoLund, Lund University, S-221 00 Lund, Sweden
}

\author{Viktor Svensson}
\affiliation{%
Division of Solid State
Physics and NanoLund, Lund University, S-221 00 Lund, Sweden
}
 \author{Martin Leijnse}
 \affiliation{%
Division of Solid State
Physics and NanoLund, Lund University, S-221 00 Lund, Sweden
}

\date{\today}

\begin{abstract}
The possibility to engineer a Kitaev chain in quantum dots coupled via superconductors has recently emerged as a promising path toward topological superconductivity and possibly nonabelian physics. Here, we show that it is possible to avoid some of the main experimental hurdles on this path by using only local proximity effect on each quantum dot in a geometry that resembles a two-dot version of the proposal in  New J. Phys. {\bf 15} 045020 (2013). There is no need for narrow superconducting couplers, additional Andreev bound states, or spatially varying magnetic fields; it suffices with spin-orbit interaction and a constant magnetic field, in combination with control of the superconducting phase to tune the relative strengths of elastic cotunneling and an effective crossed-Andreev-reflection-like process generated by higher-order tunneling. We use a realistic spinful, interacting model and show that high-quality Majorana bound states can be generated already in a double quantum dot.
\end{abstract}

\maketitle


\section{\label{sec:introduction} Introduction}
Efforts to engineer a topological superconducting phase hosting Majorana bound states (MBSs)~\cite{Wilczek2009, Alicea_RPP2012, Leijnse_Review2012, AguadoReview, BeenakkerReview_20, flensberg2021engineered, Yazdani2023} have led to encouraging experimental progress (see Refs.~\cite{Mourik_science2012, deng2012anomalous, finck2013anomalous, NadjPerge2014, deng2016majorana, Nichele_PRL2017, lutchyn2018majorana, Fornieri2019, Ren2019, Vaitiekenas2020, Aghaee2023} for some examples). However, it has also become clear that the imperfections and defects that are inevitable in real materials may lead to the emergence of nontopological Andreev bound states (ABSs) that can mimic many experimental signatures of MBSs~\cite{Prada_PRB2012,Kells_PRB12,Liu2012,Roy_PRB2013,Liu2017,Moore_PRB18,reeg2018zero,Awoga_PRL2019,Vuik_SciPost19,Pan_PRR20,Prada_review,hess2021local}. One way to avoid the problems associated with disorder is to build up a discrete Kitaev chain~\cite{Kitaev_2001} from quantum dots (QDs) coupled via superconducting segments~\cite{Sau_NatComm2012, leijnse_parity_2012}. Reaching a true topological phase requires long chains, but it was shown in Ref.~\cite{leijnse_parity_2012} that states which share all the properties of topological MBSs appear already in a minimal (two-site) Kitaev chain, based on two spin-polarized QDs coupled via crossed Andreev reflection and elastic cotunneling mediated by a single narrow superconductor. The catch is that these states, which were called poor man's Majoranas (PMMs), only appear at finetuned sweet spots in parameter space, namely when both QD levels align with the chemical potential of the superconductor and the amplitudes for crossed Andreev reflection and elastic cotunneling are equal. Reference \cite{leijnse_parity_2012} suggested tuning the ratio of crossed Andreev reflection and elastic cotunneling amplitudes via a spatially inhomogeneous magnetic field. 

The realization of the minimal Kitaev chain and PMMs suffered from difficulties associated with both controlling the inhomogeneous magnetic field and with reaching sufficiently large crossed Andreev reflection and elastic cotunneling amplitudes, which ultimately determine the gap to excited states. Both these problems were solved by an elegant proposal to couple the QDs via an ABS in the superconducting region~\cite{PhysRevLett.129.267701}. This can lead to much stronger crossed Andreev reflection and elastic cotunneling and furthermore allows tuning their ratio by controlling the energy of the ABS because of an interference effect. Furthermore, it was shown in Ref.~\cite{tsintzis_creating_2022} that high-quality PMMs also survive in the regime of large QD-ABS couplings, realistic Zeeman fields, and strong Coulomb interactions. The recent experimental breakthrough reported in Ref.~\cite{Dvir2023} (see also Refs.~\cite{Wang2022, Wang2022a, Bordin2022, Bordin2023}) showed transport spectroscopy data that are fully consistent with PMMs. However, coupling the QDs via an ABS might not be possible in all material platforms, restricts the maximum possible gap to excited states and is associated with difficulties and extensive tuning, in particular when going to longer QD chains or when coupling several PMM systems to, for example, study nonabelian physics.

In this work, we investigate a different way to engineer a Kitaev chain which entirely eliminates the need to couple the QDs via a superconductor. Instead, each QD only has local coupling to a superconductor, while the different QDs are directly tunnel coupled to each other. An effective nonlocal pairing amplitude, needed to simulate the Kitaev chain, appears due to higher order tunneling where local Andreev reflection is followed by tunneling between the QDs. This geometry was originally introduced in Ref.~\cite{fulga_adaptive_2013}, which we extend by including the effects of intra- and inter-QD Coulomb interactions and, importantly, show that if the two superconductors are connected in a loop, the superconducting phase difference controls the amplitude of the effective nonlocal superconducting pairing. This allows us to reach the sweet spot where PMMs appear already for two QDs, which is our focus.

While we are considering the case with both QDs coupled to a superconductor, a parallel work shows that PMMs can be created by alternating normal and superconducting QDs~\cite{milesKitaevChainAlternating2023}. Tuning to the sweet spot is then achieved by either controlling the amplitude of the induced superconductivity or by tuning the strength or direction of the Zeeman field.

The paper is organized as follows. Section~\ref{sec:model} introduces the model of two interacting spinful QDs, each coupled to a superconductor. Then, Sec.~\ref{sec:noninteracting} analyses the noninteracting limit, where we obtain analytic conditions for reaching a PMM sweet spot. This is followed by considering both intra- and inter-QD Coulomb interactions in Sec.~\ref{sec:interactions}, which depending on the details can either make it harder or easier to find PMM sweet spots. Finally, we summarize and conclude in Sec.~\ref{sec:conclusions}.
\begin{figure}
    \centering
    \includegraphics[width=\linewidth]{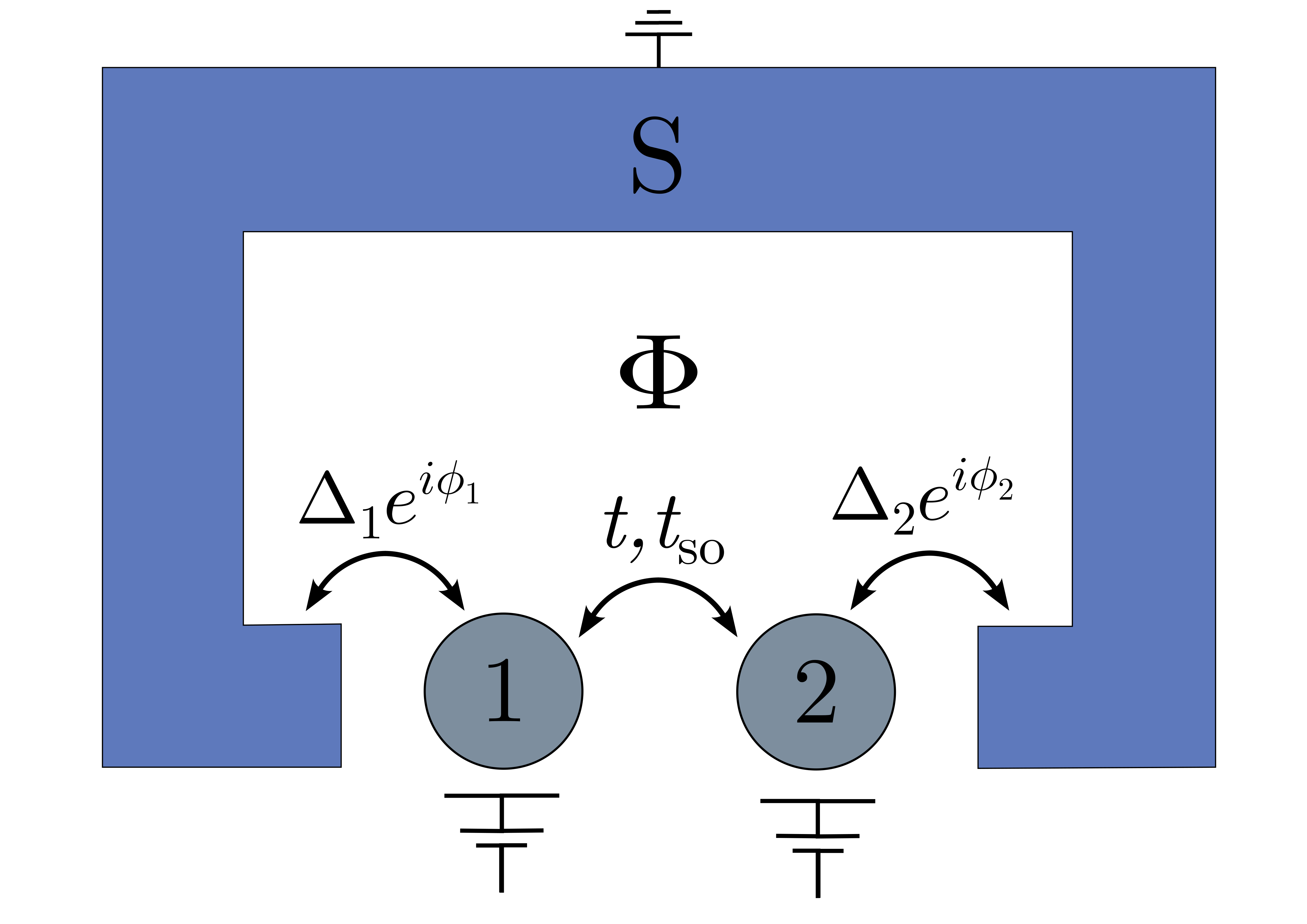}
    \caption{Setup consisting of two spinful QDs (1,2) with superconductivity induced by local tunneling to a bulk superconductor. The magnetic flux $\Phi$ controls the phase difference between the induced superconducting pairing amplitudes, $\delta\phi=\phi_1 - \phi_2$. The QDs are coupled by both spin-conserving tunneling $t$ and (spin-orbit induced) spin-flip tunneling $t_\mathrm{so}$.}
    \label{fig:model}
\end{figure}
\section{\label{sec:model} Locally proximitized double QD model}
We consider a double QD where each QD is locally proximitized by a bulk superconductor. A magnetic flux through the loop controls the phase difference between the induced pairing amplitudes of the QDs, see \cref{fig:model}. Furthermore, we consider a fully interacting system with local (intra-QD) and nonlocal (inter-QD) Coulomb interactions. Note that in other minimal Kitaev chain platforms, nonlocal Coulomb interactions are expected to be small due to the screening by the intermediate superconductor~\cite{leijnse_parity_2012}. However, we have to take such interactions into account.

The system is modeled with the Hamiltonian
\begin{equation}\label{eq:modelham}
\begin{aligned}
        H &= \sum_j H_j + t\sum_{\sigma} ( d_{1\sigma}^\dagger  d_{2\sigma} + \text{H.c.})\\  
    &+t_\mathrm{so} ( d_{1\downarrow}^\dagger d_{2\uparrow} - d_{1\uparrow}^\dagger  d_{2\downarrow} + \text{H.c.}) + U_{nl}N_1 N_2,
    \end{aligned}
\end{equation}
where $H_j$ is the Hamiltonian of QD $j$,
\begin{equation}\label{eq:QDham}
\begin{aligned}
    H_j &= \sum_{\sigma} (\varepsilon_j + \eta_\sigma V_{z,j})n_{j\sigma} \\
    &+ (\Delta_je^{i\phi_j}  d_{j\uparrow}^\dagger  d_{j\downarrow}^\dagger + \text{H.c.}) + U_{l,j} n_{j\uparrow}n_{j\downarrow}.
\end{aligned}
\end{equation}
In \cref{eq:modelham,eq:QDham}, $d_{j\sigma}$ annihilates a spin-$\sigma$ electron on QD $j=1,2$ with single-particle energy $\varepsilon_j$ relative to the chemical potential of the superconductors. We consider a single orbital on each QD and denote the total occupation on the $j$th QD with $N_j=n_{j\uparrow} + n_{j\downarrow}$, where $n_{j\sigma} = d_{j\sigma}^\dagger d_{j\sigma}$. A magnetic field induces a Zeeman splitting of $2V_{z,j}$ between the spin states on each QD. Furthermore, $\eta_{\uparrow(\downarrow)} = \mp 1$ such that the spin-up state is energetically favorable. The amplitudes for spin-conserving and spin-flip tunneling, which we choose to be real and positive, are given by $t$ and $t_\mathrm{so}$, respectively. The spin-flip tunneling results from spin-orbit interactions (we take the spin-orbit field perpendicular to the Zeeman field~\cite{stepanenko_singlet-triplet_2012}). Furthermore, we include proximity-induced superconductivity within the infinite-gap approximation with pairing terms of amplitude $\Delta_je^{i\phi_j}$, where $\phi_j$ is the phase of the superconductor proximitizing QD $j$~\cite{bauer_spectral_2007,meng_self-consistent_2009,meden_andersonjosephson_2019}. A magnetic flux $\Phi$ controls the superconducting phase difference $\delta\phi = \phi_1 - \phi_2$. Finally, we include local Coulomb interactions $U_{l,j}$ on each QD and nonlocal interactions $U_{nl}$ between the QDs.

For simplicity, we consider $V_{z,j}, \Delta_j$, and $U_{l,j}$ to be site-independent and drop their site index in the rest of the paper. However, we have verified that these assumptions do not qualitatively affect our results. Unless otherwise stated, we use $t=\Delta/2$ and $t_\mathrm{so}=t/5$.

\begin{figure}
    \centering
    \includegraphics[width=\linewidth]{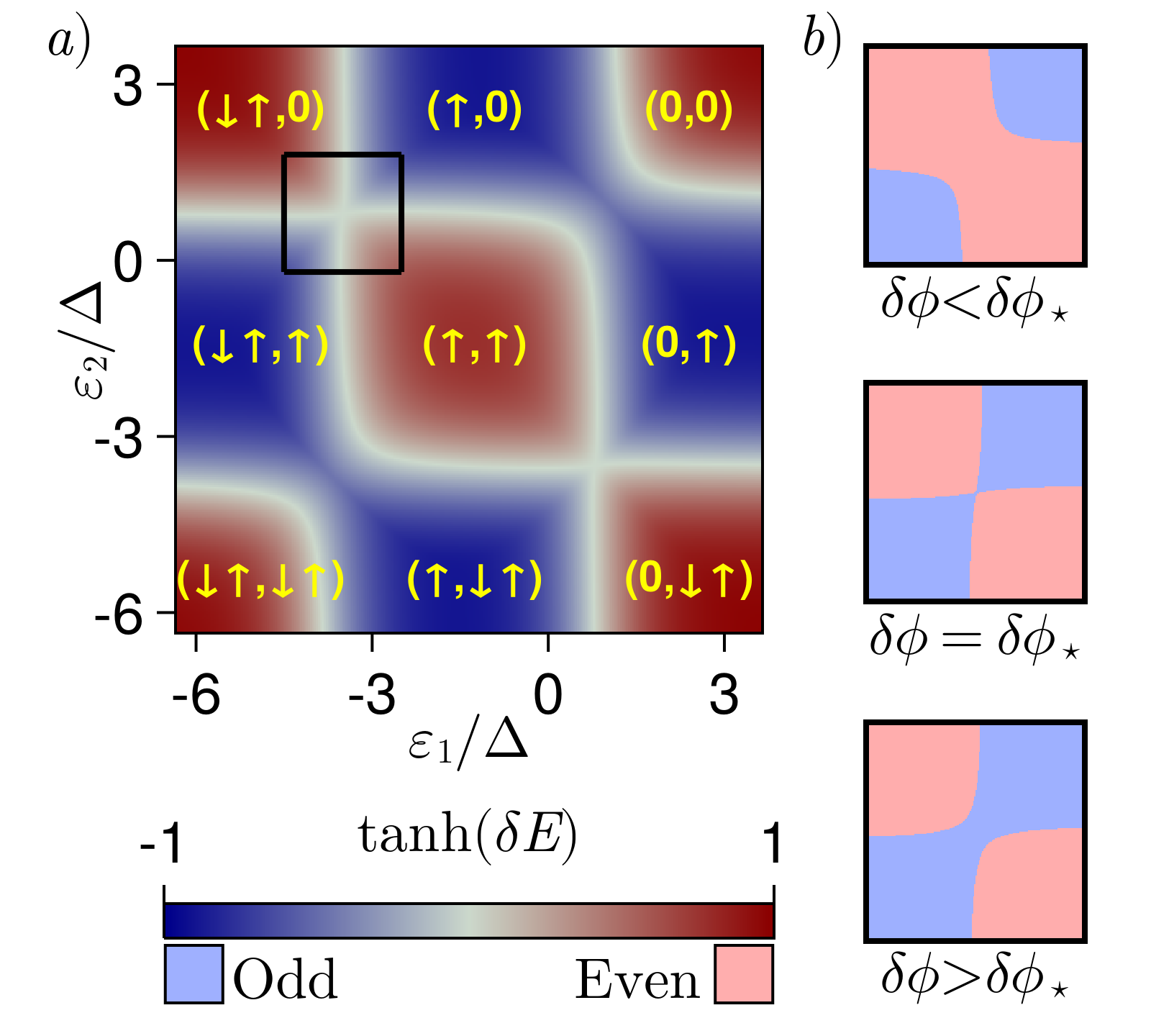}
    \caption{Calculated stability diagrams of the locally proximitized double QD using $V_z=1.25\Delta, U_l=2.5\Delta$, and $U_{nl}=0.1\Delta$. a) $\tanh{(\delta E)}$ as a function of the QD energy levels using the sweet spot phase difference $\delta\phi=\delta\phi_\star \approx 0.66 \pi$, where $\delta E = E_\mathrm{odd} - E_\mathrm{even}$ is the energy difference between the odd and even parity ground states. The black box indicates the location of the anti-parallel spin configuration, which is our focus. The labels indicate the (QD1, QD2) ground-state occupation for $\Delta = t = t_\mathrm{so} = 0$. b) Parity of the ground state as a function of the QD energy levels in the anti-parallel spin configuration, using the same parameters as in a), except the superconducting phase difference which is varied with $\pm \pi/2$.}
    \label{fig:charge_stability_phase}
\end{figure}

The basic physics of the model is shown in Fig.~\ref{fig:charge_stability_phase}. The energy difference $\delta E$ between the lowest-energy odd and even states is plotted in Fig.~\ref{fig:charge_stability_phase}(a) as a function of $\varepsilon_1$ and $\varepsilon_2$. The labels indicate the (QD1, QD2) ground-state occupations for vanishing $\Delta, t$ and $t_\mathrm{so}$. Ground-state changes are accompanied by even-odd degeneracies (narrow white regions, $\delta E = 0$). Sweet spots are found where two such degeneracy lines cross. We will investigate the conditions for such crossings below. Figure~\ref{fig:charge_stability_phase}(b) shows the ground state parity (even or odd) around the upper left crossing in (a) and demonstrates that $\delta \phi$ can be used to tune between a crossing (a sweet spot) and an avoided crossing (not a sweet spot).

\section{Noninteracting limit\label{sec:noninteracting}}
In this section, we analyze the Hamiltonian in \cref{eq:modelham} with $U_{l}=U_{nl}=0$. This allows for analytical results that serve as a starting point for the interacting case discussed in \cref{sec:interactions}. We begin by showing how an effective Kitaev chain description emerges by considering the "Kitaev limit" $V_z,\Delta \gg t, t_\mathrm{so}$. Then we discuss how the system can be tuned to a PMM sweet spot. The main conclusion is that control of the QD energy levels and the superconducting phase difference is sufficient to reach the sweet spot. Such a tuning procedure requires a Zeeman energy that is larger than the induced gap, but not by too much.

\subsection{The Kitaev limit}
Following Ref.~\cite{fulga_adaptive_2013}, we explain how the Hamiltonian in \cref{eq:modelham} with $U_{l}=U_{nl}=0$ and $V_z, \Delta \gg t, t_\mathrm{so}$ maps to the minimal Kitaev chain
\begin{equation}\label{eq:kitaev}
    H_{K} = \varepsilon_{K,1} n_{K,1} + \varepsilon_{K,2}n_{K,2} + (t_K c_2^\dagger c_1 + \Delta_K c_2^\dagger c_1^\dagger + \text{H.c.}).
\end{equation}
Here, $n_{K,j}=c_j^\dagger c_j$, where $c_j$ annihilates a particle with energy $\varepsilon_{K,j}$ at site $j=1,2$ in the minimal Kitaev chain. By performing an appropriate gauge transformation, we can (in the two-site Kitaev model) choose the tunneling amplitude $t_K$ and the amplitude of the $p$-wave pairing $\Delta_K$ to be real and positive.

We first perform a Boguliobov transformation of each QD. The BdG-eigenstates of $H_j$ become 
\begin{align}
    a_j &= \frac{e^{i\phi_j/2}}{\sqrt{2\beta_j}}\left(\sqrt{\beta_j - \varepsilon_j}\ d_{j\downarrow}^\dagger + e^{-i\phi_j}\sqrt{\beta_j + \varepsilon_j}\ d_{j\uparrow}\right),\label{eq:aparticle}\\
    b_j &= \frac{e^{i\phi_j/2}}{\sqrt{2\beta_j}}\left(\sqrt{\beta_j - \varepsilon_j}\ d_{j\uparrow}^\dagger - e^{-i\phi_j}\sqrt{\beta_j + \varepsilon_j}\ d_{j\downarrow}\right), \label{eq:bparticle}
\end{align} 
with energies
\begin{align}
    E_{a_j} &= \beta_j - V_z,\label{eq:aenergy}\\
    E_{b_j} &= \beta_j + V_z,
\end{align}
where $\beta_j = \sqrt{\varepsilon_j^2 + \Delta^2}$ and $j=1,2$.

By inverting \cref{eq:aparticle,eq:bparticle}, the full Hamiltonian in \cref{eq:modelham} can be described in terms of the $a_j$ and $b_j$ operators. However, for $V_z\gg t,t_\mathrm{so}$, a low energy Hamiltonian in terms of $a_j$-particles can be found by projection onto the states with zero $b_j$-particles. 
Furthermore, we map $a_2 \rightarrow c_2$ but $a_1 \rightarrow c_1^\dagger$ (we explain why in \cref{sec:tuning}). The resulting Hamiltonian maps directly onto the Kitaev chain in \cref{eq:kitaev}, where the effective Kitaev parameters ($\varepsilon_{K,j}, t_K$ and $\Delta_K$) depend on the physical parameters in \cref{eq:modelham}, see below.
However, we want to emphasize that a nonlocal pairing amplitude $\Delta_K$ can be generated with only local proximity effect. The effective nonlocal pairing is generated by a third-order process, consisting of a local Andreev reflection onto a QD followed by one of the electrons tunneling to the other QD. Depending on how the QD energy levels are tuned, the electron either conserves or flips its spin during the tunneling process, see below.

\subsection{Tuning to the sweet spot\label{sec:tuning}}
In the minimal Kitaev chain, one PMM localized entirely on each QD appears when tuning the system to the sweet spot $\varepsilon_{K,1}=\varepsilon_{K,2}=0$ and $t_K=\Delta_K$~\cite{leijnse_parity_2012}. To fulfill the first of these conditions, the $c_j$-particles need to have zero energy (meaning that the corresponding state is aligned with the chemical potential of the bulk superconductor). Note that this is equivalent with $E_{a_j}=0$, i.e.,
\begin{equation}\label{eq:zeroenergycond}
    \varepsilon_{K,j} = 0 \implies \sqrt{\varepsilon_j^2 + \Delta^2} = V_z,
\end{equation}
see \cref{eq:aenergy}. Therefore, in the noninteracting model, the QD energy levels must be tuned such that
\begin{equation}\label{eq:mu0}
    \varepsilon_j = \pm \sqrt{V_z^2 - \Delta^2} \equiv \pm \varepsilon_0.
\end{equation}

The solutions to $\varepsilon_j$ in \cref{eq:mu0} can most easily be understood in the limit $\Delta\rightarrow 0$ when the many-body eigenstates of $H_j$ have definite charge and
\begin{equation}\label{eq:ajlimit}
    a_j^\dagger \sim \begin{cases}
        d_{j\uparrow}^\dagger, &\text{if}\ \varepsilon_j > 0,\\
        d_{j\downarrow} &\text{otherwise},
    \end{cases}
\end{equation}
where we have dropped phase factors. \Cref{eq:ajlimit} tells us that, for the positive (negative) solution of $\varepsilon_j$ in \cref{eq:mu0}, one can add and remove a spin up (spin down) particle from QD $j$ from the ground state without energy cost. Therefore, we refer to the choice of tuning the two QD levels to the same signs and the opposite signs in \cref{eq:mu0} as parallel and anti-parallel spin configurations, respectively. For a non-zero $\Delta$, however, the excitations are instead given by \cref{eq:aparticle} and consist of superpositions of particles and holes with opposite spins. Even though referring to the spins being anti-parallel or parallel is not entirely accurate in the general case, we use this terminology here.

In the anti-parallel spin configuration, the effective nonlocal pairing is generated by Andreev reflection followed by spin-conserving tunneling, while the effective $t_K$ corresponds to spin-orbit induced spin-flip tunneling. Parallel spins, on the other hand, require a spin-flip tunneling process along with Andreev reflection to generate an effective $\Delta_K$, while the hopping term corresponds to spin-conserving tunneling. The anti-parallel spin configuration, therefore, amplifies the crossed-Andreev-reflection-like process compared to parallel spins (if $t>t_\mathrm{so}$), which we will see is beneficial to create PMMs.

In the noninteracting case, anti-parallel spins mean that we tune the QD energy levels to $\varepsilon_2 = -\varepsilon_1 = \varepsilon_0$, corresponding to the upper left corner in the stability diagram in \cref{fig:charge_stability_phase}(a) (although this plot includes interactions). This choice of tuning the QD energy levels results in $a_1$ becoming hole-like, see \cref{eq:ajlimit}, which explains our mapping $a_1\rightarrow c_1^\dagger$ above. In the anti-parallel spin configuration, the effective Kitaev parameters in \cref{eq:kitaev} become
\begin{align}
    \varepsilon_{K,1}&=\varepsilon_{K,2} = 0, \label{eq:eff_mu}\\
    t_K &= t_\mathrm{so}\left|\cos{\frac{\delta\phi}{2}} + i\frac{\varepsilon_0}{V_z} \sin{\frac{\delta\phi}{2}}\right| ,\label{eq:eff_t}\\
    \Delta_K  &= \frac{t\Delta}{V_z} \left|\sin{\frac{\delta\phi}{2}}\right|\label{eq:eff_Delta},
\end{align}
where we have taken the absolute values of the right-hand sides in \cref{eq:eff_t,eq:eff_Delta} since the phases can be gauged away. Note that $\Delta_K$ decreases with increasing $V_z$ unless we also increase $\Delta$, so we should consider the limit $V_z,\Delta \gg t, t_\mathrm{so}$. Furthermore, \cref{eq:eff_Delta} reinforces the intuitive picture where the nonlocal pairing is generated by local Andreev reflection followed by spin-conserving tunneling between the QDs.

We also require $t_K = \Delta_K$ to reach the sweet spot. From \cref{eq:eff_t,eq:eff_Delta}, we note that $t_K$ ($\Delta_K$) decreases (increases) monotonically with the superconducting phase difference $\delta\phi$ as it is varied from 0 to $\pi$, taking values within the ranges
\begin{align}
    t_K &\in \left[\frac{t_\mathrm{so}\varepsilon_0}{V_z}, t_\mathrm{so}\right],\label{eq:ranges1}\\
    \Delta_K&\in \left[0, \frac{t\Delta}{V_z}\right]\label{eq:ranges2}.
\end{align}
Therefore, if
\begin{equation}\label{eq:ineqfortuning}
    t_\mathrm{so}\varepsilon_0\leq t\Delta,
\end{equation}
the superconducting phase difference can be used to tune to the sweet spot. Otherwise, $t_K$ is always larger than $\Delta_K$. For a given $\Delta$, \cref{eq:ineqfortuning,eq:mu0} imply an upper bound for the Zeeman energy. Furthermore, according to \cref{eq:zeroenergycond}, the Zeeman energy must be larger than $\Delta$. $V_z$ is therefore bounded by
\begin{equation}\label{eq:Vzconds}
   \Delta\leq V_z \leq \Delta \sqrt{1+\left(\frac{t}{t_\mathrm{so}}\right)^2}\equiv V_{z,\mathrm{max}}.
\end{equation}

If the inequality in \cref{eq:Vzconds} is fulfilled, the tuning procedure only involves the QD energy levels and the superconducting phase difference. Firstly, the QD levels can be tuned separately such that $\varepsilon_{K,1/2} = 0$. Then, tuning the superconducting phase difference to a sweet spot value $\delta\phi_\star$ is sufficient to achieve $\Delta_K=t_K$ and end up at the PMM sweet spot. The stability diagrams in \cref{fig:charge_stability_phase}(b) illustrate the tuning procedure. When increasing the superconducting phase difference, we can tune from having an even-parity dominated anti-crossing at $\delta\phi < \delta\phi_\star$ ($t_K > \Delta_K$), to the PMM sweet spot at $\delta\phi = \delta\phi_\star$ ($t_K = \Delta_K$), to an odd-parity dominated anti-crossing ($t_K < \Delta_K$) for even larger phase differences.

To the extent that the PMMs are separated, the energy degeneracy is protected to first order against local perturbations. This includes perturbations of the dot levels and the superconducting phases~\footnote{A perturbation in the superconducting phase effectively perturbs the local pairing by $d\Delta = i\Delta d\phi$. The size of this perturbation is proportional to $\Delta$, so the protection against phase fluctuations does not necessarily improve in the Kitaev limit $V_z \rightarrow \infty$ and $\Delta \rightarrow \infty$.}.


\subsection{Energy gap and spin configuration}
At the sweet spot in the minimal Kitaev chain, the degenerate ground states are separated by a gap $E_g=2t_K=2\Delta_K$ to the next excited states~\cite{leijnse_parity_2012}. To detect PMMs we need the gap to be much larger than the thermal energy. Furthermore, both qubit coherence times and the time-scale requirements for nonabelian operations will benefit from a large gap.

To provide an estimate of the resulting energy gap in our system, consider the situation when $V_z = V_{z,\mathrm{max}}$. Then, $t_K = \Delta_K$ at $\delta\phi=\pi$, and we can obtain the energy gap by substituting $V_z$ by $V_{z,\mathrm{max}}$ in \cref{eq:ranges2} and doubling the result:
\begin{equation}\label{eq:gap}
    E_g = \frac{2t\Delta}{V_{z,\mathrm{max}}} = \frac{2t}{\sqrt{1+(t/t_\mathrm{so})^2}}.
\end{equation}
Compared with numerical results at the sweet spot in the full model, we find that \cref{eq:gap} provides a good estimate of the gap in general.

If arranging the spins parallel instead, i.e., by tuning both QD energy levels to $\varepsilon_0$ (or both to $-\varepsilon_0$), the bound on $V_z$ in \cref{eq:Vzconds} and the energy gap in \cref{eq:gap} are acquired by mapping $t\longleftrightarrow t_\mathrm{so}$. Anti-parallel spins, therefore, allow for larger Zeeman energies than parallel spins (if $t>t_\mathrm{so}$), while the energy gap is the same for both configurations. Since we have not found any advantages of the parallel spin configuration when $t>t_\mathrm{so}$, we focus on the anti-parallel case. However, if $t<t_\mathrm{so}$, the parallel spin configuration is superior and we get almost identical results as the anti-parallel spins if $t_\mathrm{so}/t \rightarrow t/t_\mathrm{so}$. 

\subsection{Relaxing the Kitaev limit\label{sec:relaxingkitaev}}
Due to the restrictions on the Zeeman energy in \cref{eq:Vzconds}, we cannot increase $V_z$ indefinitely without also increasing $\Delta$, which in turn is bounded by the gap of the parent superconductor. Furthermore, superconductivity might get quenched by the large magnetic field even before the upper bound in \cref{eq:Vzconds}. Another option to approach the Kitaev limit is to decrease the coupling between the QDs. However, according to \cref{eq:gap}, a smaller coupling (i.e., smaller $t$ and $t_\mathrm{so}$, but fixed $t_\mathrm{so}/t$) results in a decreased energy gap. Practically, there is hence a trade-off between approaching the Kitaev limit (small coupling) on one hand and a large energy gap (large coupling) on the other.

It was shown in Ref.~\cite{tsintzis_creating_2022} that a large Zeeman energy is necessary to get well-separated PMMs, but our system is rather different and the precise relation between Zeeman energy and PMM separation might be different. Furthermore, we don't expect the expressions for the effective Kitaev parameters in \cref{eq:aenergy,eq:eff_t,eq:eff_Delta} to be correct in general, since they were derived in the limit $V_z \gg t,t_\mathrm{so}$. Therefore, when relaxing the Kitaev limit, it becomes difficult to predict where a possible sweet spot is located. We will below introduce a PMM quality measure that we can numerically optimize to locate the best sweet spot and evaluate its quality.

\section{Interactions and Majorana quality\label{sec:interactions}}
Having understood how the noninteracting model can be described by an effective minimal Kitaev chain in the Kitaev limit, we now discuss the influence of interactions. Then we define a PMM quality measure and discuss how we locate PMM sweet spots by numerical optimization. Finally, in \cref{sec:fullsystem}, we use the optimization procedure to find experimentally relevant parameter regimes with high-quality PMMs.

\subsection{Local Coulomb interactions\label{sec:intraQD}}
With local Coulomb interactions, we can still intuitively understand the system as a minimal Kitaev chain in the Kitaev limit. However, local Coulomb interactions renormalize the effective Kitaev parameters, resulting in a new sweet spot location. By considering the many-body picture of $H_j$ in \cref{eq:QDham}, we can get a new estimate for the $\varepsilon_{K,1} = \varepsilon_{K,2} = 0$ sweet spot condition, providing an initial guess for the optimization procedure. We seek solutions of $\varepsilon_j$ such that $H_j$ has degenerate ground states. The ground state of $H_j$ is either the pure spin-up state or the low-energy BCS-singlet. Solving for their degeneracy modifies the condition in \cref{eq:mu0} to
\begin{equation}\label{eq:mu0_interacting}
    \varepsilon_j = \pm \sqrt{\left(V_z + \frac{U_{l}}{2}\right)^2 - \Delta^2} - \frac{U_{l}}{2}.
\end{equation}
To obtain the anti-parallel spin configuration, the QD levels are tuned to have opposite signs in \cref{eq:mu0_interacting}. Furthermore, the lower bound for $V_z$ becomes $V_z + U_{l}/2 \geq \Delta$. Therefore, if $U_{l}\geq 2\Delta$, there is no lower bound on $V_z$ to achieve $\varepsilon_{K,1}=\varepsilon_{K,2}=0$.

The local Coulomb interactions also affect the effective Kitaev tunneling $t_K$ and $p$-wave pairing $\Delta_K$, and we do not find simple, analytical expressions for them in the general case. Therefore, we cannot find an upper bound for $V_z$ as in \cref{eq:Vzconds}. As a rough estimate of the upper bound, we can consider the mean-field correction to the Zeeman energy due to the local Coulomb interaction $V_z \rightarrow V_z + U_{l}/2$ and apply it to the noninteracting bound in \cref{eq:Vzconds}. The estimated bound then becomes 
\begin{equation}\label{eq:Vzconds_interactions}
   \Delta\leq V_z + \frac{U_{l}}{2}\leq V_{z,\mathrm{max}}.
\end{equation}
\subsection{Nonlocal Coulomb interactions\label{sec:intersitekitaev}}
To gain intuition about the effect of nonlocal interactions, we study the interacting minimal Kitaev chain 
\begin{equation}\label{eq:interactingkitaev}
    H_{K,\mathrm{int}}= H_K + U_Kn_{K,1}n_{K,2},
\end{equation}
where $U_K$ is an intersite interaction. We now search for a PMM sweet spot in the interacting minimal Kitaev chain, where the odd and even parity ground states are degenerate, and one PMM is localized on each QD. As was shown in Ref.~\cite{brunetti_anomalous_2013}, such a sweet spot can be found at the point
\begin{align}
      \varepsilon_{K,1} &=  \varepsilon_{K,2} = -\frac{U_K}{2}, \label{eq:ssinteracting1}\\
    \Delta_K &= t_K + \frac{U_K}{2} \label{eq:ssinteracting2}.
\end{align}
At the point in \cref{eq:ssinteracting1,eq:ssinteracting2}, the odd and even parity ground states of $H_{K,\mathrm{int}}$ are degenerate, and all eigenstates are the same as at the sweet spot in the noninteracting case, analyzed in detail in Ref.~\cite{leijnse_parity_2012}. Therefore, the same PMMs also appear in the interacting case. However, the PMMs in the interacting minimal Kitaev chain only map between the odd and even states in the ground state sector, not the whole spectrum (the excited states are not degenerate). Therefore, the PMMs in the interacting minimal Kitaev chain are weak Majoranas~\cite{alicea_topological_2016}.

In the presence of nonlocal interactions $U_{nl}$ in the full model, the sweet spot conditions are given by \cref{eq:ssinteracting1,eq:ssinteracting2}, where $U_K$ is an effective intersite interaction proportional to $U_{nl}$. Furthermore, the nonlocal interactions renormalize the other effective Kitaev parameters. Therefore, we must rely on a numerical optimization procedure to locate PMM sweet spots when interactions are present.

\subsection{Majorana polarization and optimizing for the PMM sweet spot\label{sec:mp_optimizing}}
To utilize an optimization procedure to locate PMM sweet spots, we need to define a PMM quality measure to optimize. In this work, we quantify PMM quality by ground state degeneracy and the Majorana polarization (MP), which provides a measure of the separation between the PMMs. A large MP is necessary to observe topologically protected nonabelian physics~\cite{tsintzis_roadmap_2023}.

Since the Hamiltonian in \cref{eq:modelham} is complex, we need to generalize previous formulations of the MP~\cite{aksenov_strong_2020,tsintzis_creating_2022}. In \cref{app:majorana_polarization}, we derive the expression
\begin{equation}\label{eq:majoranapolarization}
    \mathrm{MP}_j = \frac{\abs{\sum_{\sigma,s} \matrixelement{e}{\gamma_{j\sigma s}}{o}^2}}{\sum_{\sigma,s} \abs{\matrixelement{e}{\gamma_{j\sigma s}}{o}^2}},
\end{equation}
where $\ket{o}$ and $\ket{e}$ are the odd and even ground states and
\begin{align}
    \gamma_{j\sigma +} &= d_{j\sigma} + d_{j\sigma}^\dagger, \\
    \gamma_{j\sigma -} &= i(d_{j\sigma} - d_{j\sigma}^\dagger)
\end{align}
are the local Majorana operators on QD $j$ with spin $\sigma$. $\mathrm{MP}_j$ takes values between 0 and 1, where 1 indicates that on QD $j$, the lowest-energy excitation is entirely Majorana-like, while 0 indicates that it is entirely fermion-like. As our system is not symmetric, it is important to ensure that both $\mathrm{MP}_1$ and $\mathrm{MP}_2$ are large, so we define the total $\mathrm{MP}$ as
\begin{equation}
    \mathrm{MP} = \frac{\mathrm{MP}_1 + \mathrm{MP}_2}{2}.
\end{equation}
At the sweet spot, we would like the $\mathrm{MP}$ to be close to one and the energy difference between the odd and even ground states close to zero. For details on the numerical optimization, see \cref{app:opt}.

\subsection{Moderate Zeeman energies \label{sec:fullsystem}}
Having separately discussed the effects of relaxing the Kitaev limit and of local and nonlocal interactions for creating PMMs in our model, we now study the system in its entirety. In particular, we want to find realistic parameter regimes such that we only need to fine-tune the QD energy levels and the phase difference to reach a PMM sweet spot. We will call such regions "PMM-compatible". To find PMM-compatible regimes, we perform the optimization procedure while varying $U_l$ and either $V_z$ or $U_{nl}$ and see what regimes yield a large MP. 

\begin{figure}
   \centering
   \includegraphics[width=\linewidth]{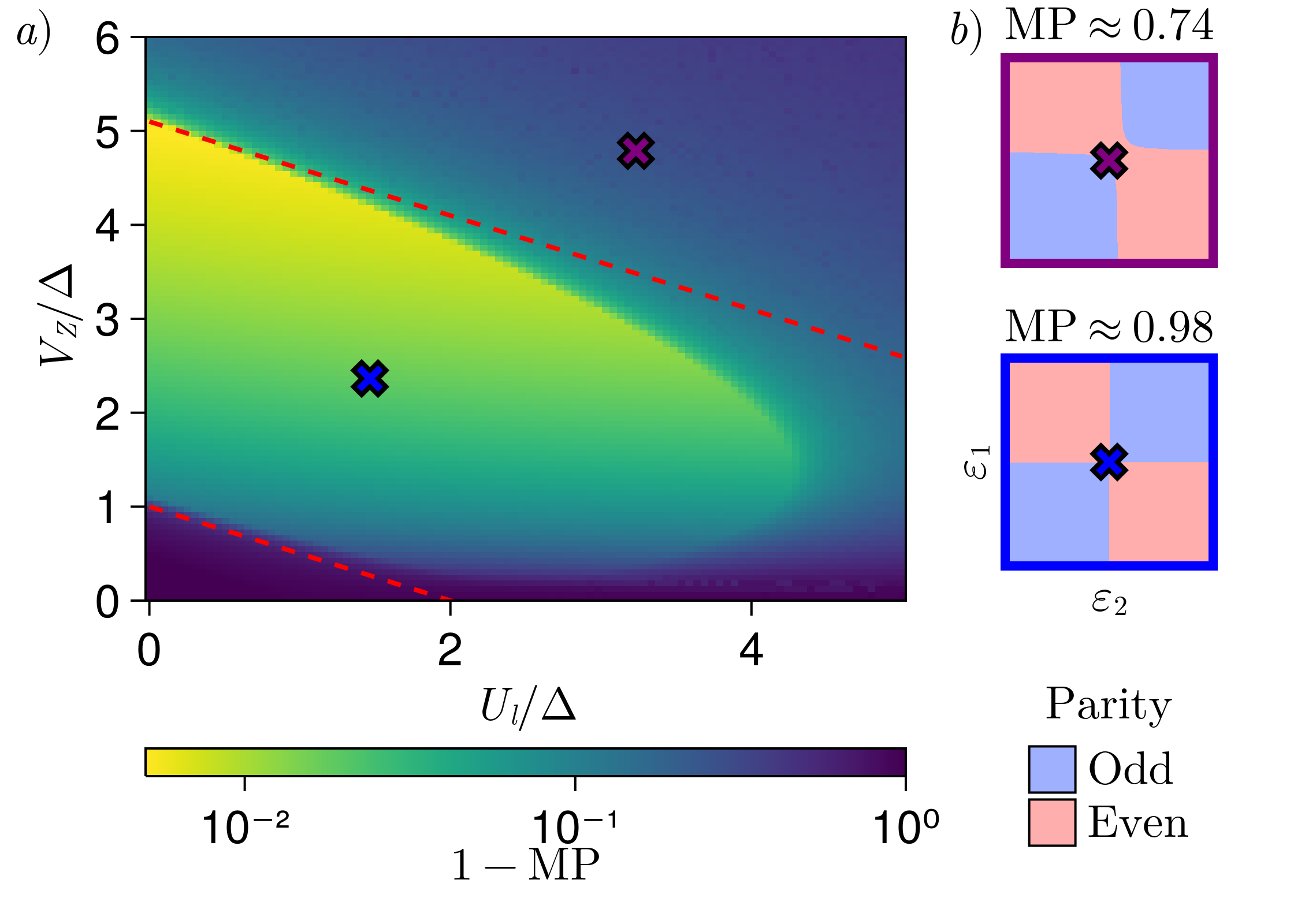}
   \caption{a) $1-\mathrm{MP}$ as a function of $U_l$ and $V_z$, using $U_{nl}=0$. At each point, $\varepsilon_1, \varepsilon_2$, and $\delta\phi$ are optimized to find a PMM sweet spot. The dashed lines represent the upper and lower bounds in \cref{eq:Vzconds_interactions}, estimating where we can tune to a PMM sweet spot. For each marking in a), a stability diagram indicating the ground state parity is shown in b). The stability diagrams are centered around the optimized QD energy levels while using the optimized phase difference. Furthermore, the MP at each marking is shown above the corresponding stability diagram.}
   \label{fig:uhplot}
\end{figure}

We first consider the case $U_{nl}=0$. Varying the Zeeman energy $V_z$ and the local Coulomb interaction $U_{l}$, we optimize $\varepsilon_1, \varepsilon_2$ and $\delta\phi$ for all combinations of $(U_{l}, V_z)$. \Cref{fig:uhplot}(a) shows the resulting heat map of $1 - \mathrm{MP}$ at the optimized points. There is a distinct PMM-compatible region in the bright yellow and green area showing that we can create high-quality PMMs with moderate Zeeman energies. The PMM-compatible region is surrounded by darker blue regions with poor MP at large $U_{l}$ and/or $V_z$ in the top right and at small $U_{l}$ and $V_z$ in the bottom left. The lower dashed line in \cref{fig:uhplot}(a) shows the lower bound in \cref{eq:Vzconds_interactions}, meaning that the $\varepsilon_{K,1} = \varepsilon_{K,2} = 0$ condition cannot be fulfilled for values of $V_z$ and $U_l$ below the line. This bound agrees well with the boundary to the PMM-compatible region.

Within the PMM-compatible region, increasing the Zeeman energy improves the MP. An increased local Coulomb interaction also improves the MP, and for $V_z < \Delta$, a non-zero $U_l$ is necessary to enter the PMM-compatible region. However, too large $U_l$ and $V_z$ both cause problems for tuning to a PMM sweet spot. We understand the problem for large $V_z$ and/or $U_l$ as the failure of fulfilling the $t_K = \Delta_K$ condition while tuning $\delta\phi$. In the Kitaev limit of the noninteracting model discussed in \cref{sec:noninteracting}, we derived a bound on $V_z$ such that if $V_z > V_{z,\mathrm{max}}$, all $\delta\phi$ yield $t_K > \Delta_K$. We also estimated the corresponding upper bound in the presence of local Coulomb interactions in \cref{eq:Vzconds_interactions}, which is included in \cref{fig:uhplot}(a) as the upper dashed line. To test if an insufficient $\Delta_K$ is the problem for large $U_{l}$, we study the purple marking in \cref{fig:uhplot}(a), lying outside the PMM-compatible region. In the corresponding stability diagram in \cref{fig:uhplot}(b), we see an anti-crossing with a dominating even ground state for all $\delta\phi$. As discussed in \cref{sec:tuning}, an even-parity-dominated anti-crossing is consistent with $t_K > \Delta_K$, indicating that the effective $\Delta_K$ is too small to find a PMM sweet spot at large $V_z$ and/or $U_l$. In contrast, the blue marking in \cref{fig:uhplot}(a) corresponds to a crossing in the stability diagram in \cref{fig:uhplot}(b), indicating a sweet spot.

\begin{figure}
   \centering
   \includegraphics[width=\linewidth]{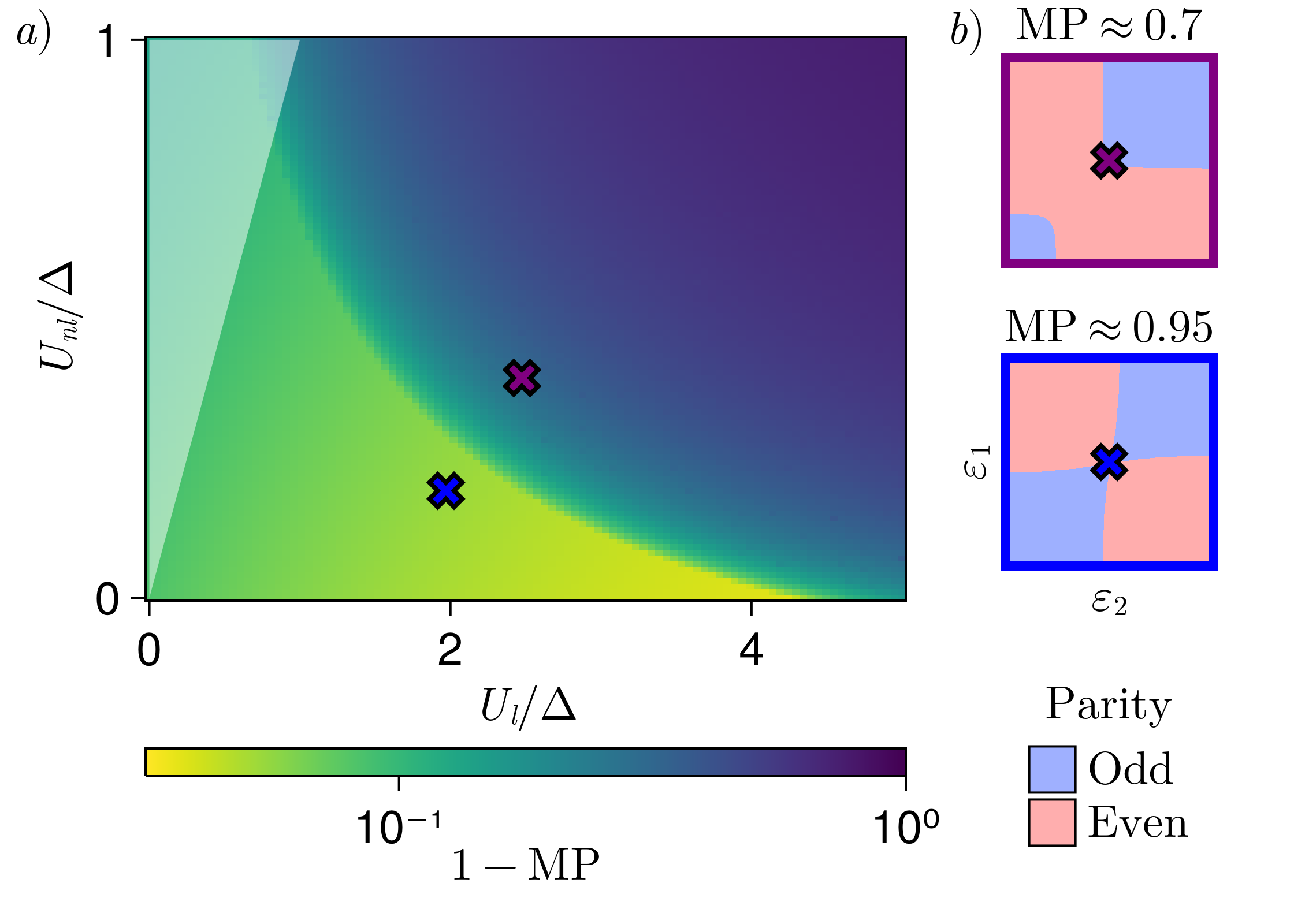}
   \caption{a) $1-\mathrm{MP}$ as a function of $U_l$ and $U_{nl}$, using $V_z=1.25\Delta$. At each point, $\varepsilon_1, \varepsilon_2$, and $\delta\phi$ are optimized to find a PMM sweet spot. Furthermore, the region $U_{nl}>U_l$ is blurred since it is unphysical in a capacitive model~\cite{vanderwielElectronTransportDouble2002}. b) The MP at the sweet spot and the stability diagram showing the ground state parity at each marking in a).}
   \label{fig:uvplot}
\end{figure}

Next, we include nonlocal interactions $U_{nl}$. In \cref{fig:uvplot}(a), we perform the optimization procedure at each point in the $(U_l, U_{nl})$-plane, fixing the Zeeman energy to $V_z=1.25\Delta$. Despite nonlocal interactions, we still find a bright, PMM-compatible region. \Cref{fig:uvplot}(b) shows the stability diagrams at the blue and purple markings in \cref{fig:uvplot}(a). Note that the stability diagram corresponding to the high-MP case (blue marking) shows a tilted crossing in contrast to the straight crossing in the high-MP sweet spot in \cref{fig:uhplot}(b). The tilted crossing is an inherent feature of sweet spots in the interacting minimal Kitaev chain and is present even when $\mathrm{MP}=1$.

In \cref{fig:uvplot}(a), it is clear that the region with large $U_l$ and/or $U_{nl}$ is associated with poor MP. We seek a simple explanation for why large $U_{nl}$ does not allow for PMM sweet spots based on the interacting Kitaev chain discussed in \cref{sec:intersitekitaev}. In the interacting minimal Kitaev chain, the sweet spot condition of equal $t_K$ and $\Delta_K$ becomes $\Delta_K = t_K + U_K/2$, where $U_K\propto U_{nl}$ in the Kitaev limit of our full system. Therefore, nonlocal interactions increase the minimum amplitude of $\Delta_K$ required to reach a sweet spot, compared to when $U_{nl}=0$. To support this reasoning, \cref{fig:uvplot}(b) shows a stability diagram also for the purple marking in \cref{fig:uvplot}(a). There, we see an even parity-dominated anti-crossing for all $\delta\phi$, indicating that, indeed, $\Delta_K < t_K + U_K/2$ for large nonlocal interactions.

In \cref{app:extra}, we provide additional data related to \cref{fig:uhplot,fig:uvplot}, consisting of heat maps of the optimized $\varepsilon_1, \varepsilon_2$, and $\delta\phi$. Furthermore, we have calculated the small energy splitting between the nearly degenerate odd and even ground states at the optimized sweet spots, as well as the energy gap to the lowest excited state (data not shown). Inside the PMM-compatible regions, the degeneracy is never broken by more than $10^{-3}\Delta$ and the gap to excited states is rather constant and approximately given by \cref{eq:gap}. Therefore, we can trust that the optimization does not improve the MP by decreasing the energy gap or by failing to have degenerate ground states.

Finally, we briefly discuss the effect of varying the tunneling amplitudes $t$ and $t_\mathrm{so}$. Firstly, increasing the tunneling amplitudes while keeping $t_\mathrm{so}/t$ fixed improves the energy gap but results in a smaller MP in the PMM-compatible region. Secondly, if $t_\mathrm{so}/t$ is increased (fixing $t$), the energy gap improves, but the PMM-compatible region becomes smaller. This behavior can be understood in the Kitaev limit in the noninteracting model, where \cref{eq:Vzconds,eq:gap} imply that an increased fraction of spin-orbit tunneling increases the gap but decreases $V_{z,\mathrm{max}}$. 

\section{Conclusions\label{sec:conclusions}}
In this work, we have proposed to engineer a minimal Kitaev chain in a double QD with local superconducting proximity effect and spin-orbit interaction. Studying a spinful and fully interacting model, we have shown that tuning to a PMM sweet spot only requires precise control of the QD energy levels and the superconducting phase difference between the two QDs. Such a tuning procedure works if the Zeeman energy and the interactions are within a "PMM-compatible" region, where the local Coulomb interactions and the Zeeman energy are not too small or too large, and the nonlocal Coulomb interaction is small. Increasing the tunneling amplitude (keeping $t_\mathrm{so}/t$ constant) increases the energy gap to excited states, but lowers the MP. Increasing $t_\mathrm{so}$ while keeping $t$ constant also increases the gap, but decreases the size of the PMM-compatible region. 

As an example of a realistic set of parameter values with a sweet spot with reasonable MP and an adequate energy gap, consider the blue marking in \cref{fig:uvplot}. There, $V_z=1.25\Delta, U_l = 2\Delta$ and $U_{nl} = 0.2\Delta$, which results in a sweet spot with $\mathrm{MP} \approx 0.95$ and $E_g \approx 0.18\Delta$. A larger $V_z$ allows for higher MP. We have compared our model to the one in Ref.~\cite{tsintzis_creating_2022}, where elastic cotunneling and crossed Andreev reflection are mediated by an intermediate QD coupled to a superconductor. We use the same values for the Zeeman energy, local Coulomb interaction, and the ratio between spin-conserving and spin-flip tunneling, but adjust the amplitude of the latter two so that the energy gap at the sweet spot is similar in both models. As long as one stays within the PMM-compatible region, the two models give similar values (within a few percent) for the $\mathrm{MP}$.

One long-term future goal for minimal Kitaev chain platforms is to create longer Kitaev chains that host topologically protected MBSs. A factor to consider when going to longer chains is that the phases of the effective Kitaev tunneling and $p$-wave pairing amplitudes cannot be removed by a gauge transformation. Since we tune the superconducting phase in the system, we will naturally acquire phases on the effective Kitaev parameters, which can suppress the resulting energy gap~\cite{Sau_NatComm2012}. However, the sweet spot condition relating the effective Kitaev tunneling and pairing amplitudes still only depends on their absolute values.  

To conclude, we emphasize that our proposal does not depend on an intermediate superconductor between the two QDs. Instead, it only relies on local superconducting proximity effect, which lowers the experimental barrier for minimal Kitaev chain platforms. The lack of an intermediate superconductor brings nonlocal interactions between the QDs into play. However, we have shown that PMMs can still be created in a fully interacting system with a simple tuning procedure. Therefore, we are hopeful that the locally proximitized double QD will provide an alternative platform for minimal Kitaev chains and next-generation experiments with PMMs.

\begin{acknowledgments}
We acknowledge enlightening discussions with Rubén Seoane Souto and Athanasios Tsintzis and funding from the European Research Council (ERC) under the European Unions Horizon 2020 research and innovation programme under Grant Agreement No. 856526, the Swedish Research Council under
Grant Agreement No. 2020-03412, and NanoLund.
\end{acknowledgments}

\appendix

\section{\label{app:majorana_polarization}Majorana polarization}
In this section, we derive an expression for the Majorana polarization (MP), and the unnormalized MP ($\mathrm{MPU}$) which provide two different measures of the separation between Majoranas. Previous formulations of the MP \cite{tsintzis_creating_2022, aksenov_strong_2020} work for models where the ground states can be chosen to be real. Since we tune the superconducting phase, we require a more general formulation. 

This derivation works by extracting the MBSs from the odd and even ground states and then defining a measure of their separation. One must keep in mind that the relative phase between the odd and even ground states influences the form of the Majorana operators that map between them. However, because of superselection, this phase is unphysical. We parametrize this freedom by $\theta$ and in the end set it to whatever maximizes the separation between the MBSs.

The Majorana operators map between the odd and even ground states as
\begin{align}\label{eq:gammadef}
    \ket{o} = e^{-i\theta}\gamma \ket{e},
\end{align}
where $\theta$ is the phase described above. Define the local Majorana operators
\begin{align}
    \gamma_{n+} &\equiv d_n + d_n^\dagger, \\
    \gamma_{n-} &\equiv i (d_n - d_n^\dagger),
\end{align}
where $n$ is a multi-index referring to, e.g., the spin and site indices. 

We assume that $\gamma$ can be expressed as a linear combination of the local Majorana operators as
\begin{equation}\label{eq:gammaansatz}
\gamma = \sum_{n,s} x_{n,s} \gamma_{n,s},
\end{equation}
where $x_{n,s}$ are real. Strictly speaking, this derivation holds only for noninteracting theories, as we impose a restriction to single-particle Majoranas in our ansatz. We assume that corrections coming from interactions are negligible. 

Combining \cref{eq:gammadef,eq:gammaansatz}, leads to
\begin{equation}
    x_{n,s} = \Re{e^{i\theta}z_{n,s}},
\end{equation}
where we have defined 
\begin{equation}
    z_{n,s} = \matrixelement{e}{\gamma_{n,s}}{o}.
\end{equation}
Choosing two different values for $\theta$, with $\Delta \theta = \pi/2$, gives two different Majorana operators. The wavefunction of the other Majorana is then given by the imaginary part as
\begin{equation}
    y_{n,s} = \Im{e^{i\theta}z_{n,s}}.
\end{equation}

The better separated these two Majoranas are from each other, the better protected the system is from perturbations. Their separation $S_R$ in a region $R$ can be measured by
\begin{equation}
    S_R(\theta) = \sum_{s,n \in R} x_{n,s}^2 - y_{n,s}^2 = \sum_{s,n \in R} \Re{e^{2i\theta}z_{n,s}^2},
\end{equation}
which depends on the phase $\theta$. We can choose this phase so that the separation is maximal, where it then takes the value
\begin{equation}
    \max_\theta S_R(\theta) = \abs{\sum_{s, n \in R } z_{n,s}^2}\equiv \mathrm{MPU}_R.
\end{equation}
$\mathrm{MPU}$ is closely connected to the MP by
\begin{equation}\label{eq:MPMPu}
    \mathrm{MP}_R = \frac{\mathrm{MPU}_R}{\sum_{s,n\in R} \abs{z_{n,s}}^2},
\end{equation}
which reduces to the MP definition in Refs.~\cite{aksenov_strong_2020,tsintzis_creating_2022} when the ground states can be chosen to be real. From \cref{eq:MPMPu}, it is clear why we refer to $\mathrm{MPU}$ as the unnormalized MP.

\section{Optimization}\label{app:opt}
By tuning $\varepsilon_1$, $\varepsilon_2$ and $\delta\phi$, we seek a sweet spot where there is an exact energy degeneracy between the odd and even ground states and well-separated Majorana modes. In principle, one can impose an exact ground state degeneracy since there is no avoided crossing between odd and even states. In practice, we use a penalty factor that ensures that $|\delta E|<10^{-3} \Delta$. 

To quantify the separation of Majorana modes during the optimization, we use the symmetric \textit{unnormalized} Majorana polarization $\mathrm{MPU} = (\mathrm{MPU}_1 + \mathrm{MPU}_2)/2$. We found that the unnormalized version is more robust than the normalized version when the energy levels are tuned independently. In the PMM-compatible region it doesn't matter which measure is used, both measures result in very similar sweet spots. 

The code to reproduce our calculations can be found at~\cite{code}.

\section{Supplementary results\label{app:extra}}
\begin{figure}
    \centering\includegraphics[width=\linewidth]{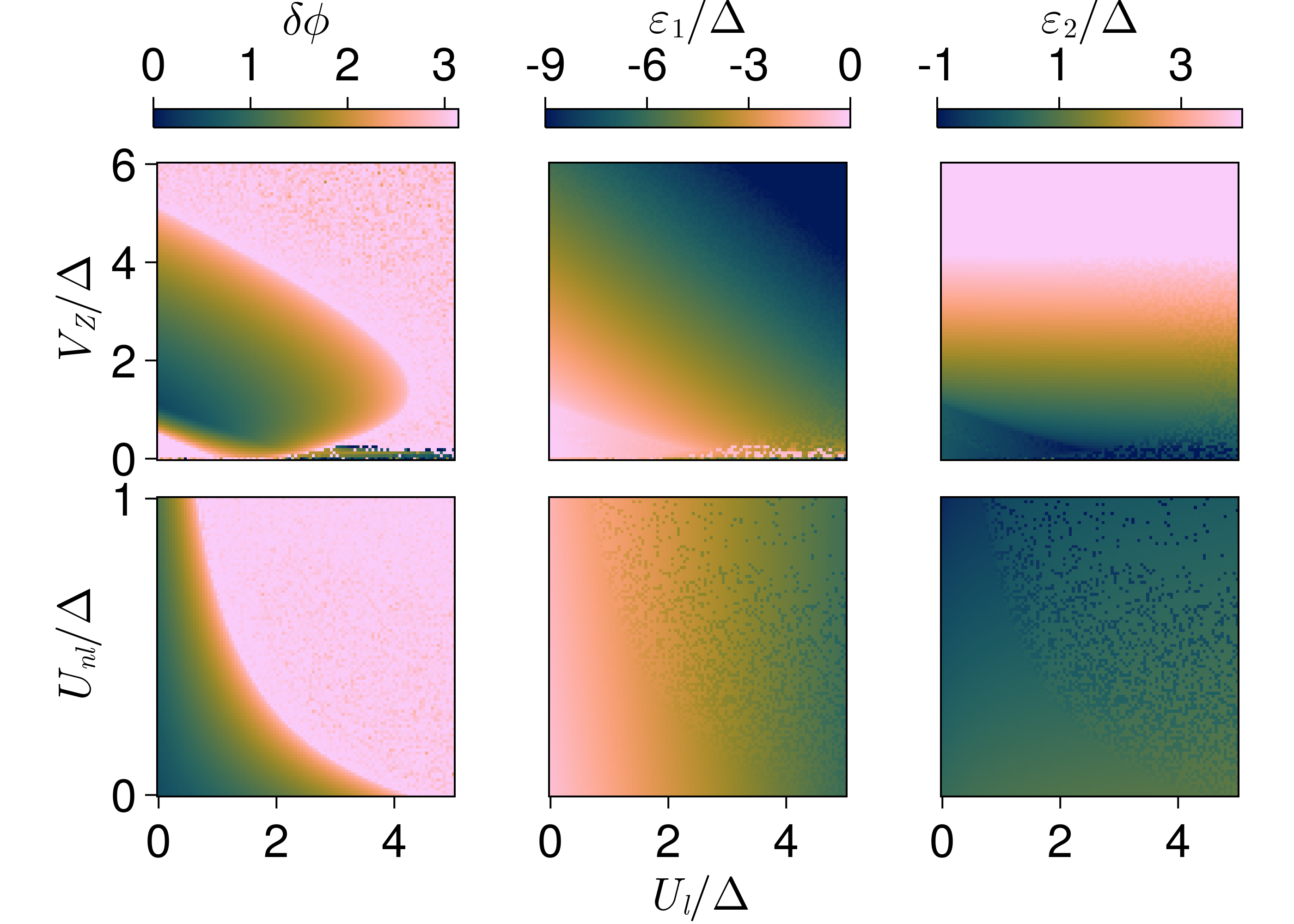}
    \caption{$\varepsilon_1, \varepsilon_2$, and $\delta\phi$ after sweet spot optimization in the $(U_l, V_z)-$ and $(U_l, U_{nl})-$planes. The values for $U_{nl}$ in the upper panels and for $V_z$ in the lower panels are the same as in \cref{fig:uhplot,fig:uvplot}.}
    \label{fig:parameter_scans}
\end{figure}

\Cref{fig:parameter_scans} shows how $\delta\phi$, $\varepsilon_1$ and $\varepsilon_2$ at the optimized sweet spot vary with the Zeeman splitting and the interactions. Note how $\delta\phi$ gradually increases when approaching the boundary to the PMM-compatible region and that $\delta\phi=\pi$ almost everywhere outside of it. From \cref{sec:tuning}, we know that $\Delta_K$ increases monotonically with $\delta\phi$ between 0 and $\pi$ in the Kitaev limit of the noninteracting model. We can hence understand the behavior of $\delta\phi$ in \cref{fig:parameter_scans} as follows. Inside the PMM-compatible region, the $\delta\phi$ which fulfills the sweet spot condition $\Delta_K = t_K + U_K/2$ gradually increases until $\delta\phi=\pi$ when hitting the boundary. Outside of the PMM-compatible region, the optimization finds $\delta\phi=\pi$ to maximize $\Delta_K$, but it still is not enough to fulfill the sweet spot condition. See the discussion in \cref{sec:fullsystem}.

\bibliography{bibliography}

\end{document}